# High power radially polarized light generated from photonic crystal segmented half-wave-plate

P.B. Phua[1,2,*], W. J. Lai[2], Y. L. Lim[1], B. S. Tan[1], R. F. Wu[1], K. S. Lai[1] and H. W. Tan[1]

[1]*DSO National Laboratories,*

*20, Science Park Drive, S118230, Republic of Singapore,*

[2]*Nanyang Technological University,*

*50, Nanyang Avenue, S 639798, Republic of Singapore*

*\* Corresponding Author: ppohboon@alum.mit.edu*

**Abstract:**

We have generated more than 100 watts of radial polarized beam from a Yb fiber laser using a photonics crystal segmented half-wave-plate. We demonstrated the high power handling capability of such a photonics crystal segmented half-wave-plate and show that it is a promising external radial polarization converter for high power Yb fiber laser used in laser cutting industry.

©2007 Optical Society of America

Radially polarized light has gained much interest recently [1-24]. This is due to its ability to be focussed tighter than a diffraction-limited beam [1], its ability to be amplified more efficiently using isotropic rod laser amplifier than plane-polarized light, [4], and its enhanced efficiency in laser cutting and machining [21, 22]. It has been shown that using radially polarized light for laser cutting is twice as efficient as compared to using circularly polarized beam [21,22]. With this motivation, M. A. Ahmed and co-workers [19] had recently demonstrated a radially polarized 3kW beam from a $CO_2$ laser with an intra-cavity resonant grating mirror. Fiber laser, on the other hand, is also a common work-horse for the laser cutting and machining industry. Therefore, the generation of radially polarized light from fiber laser will be interesting and useful for the same reason. Although Ref. [18], had shown an intra-cavity radially polarized beam generation from fiber laser using free-space dual conical prism, an external polarization conversion method of fiber laser is still a preferred approach due to its robustness and ease of implementation. In our recent publications [14, 23, 24], we have proposed various methods that can perform this external polarization conversion. One key advantage for all these methods is their ability to handle high laser power which is mandatory for laser cutting. In Ref. [14], we mimic optical activity using linear birefringence to generate radial polarized light. It involves a spirally varying birefringent retarder sandwiched between two orthogonally-oriented quarter-wave plates. In Ref. [23], we utilize Goos-Hanchen shift of Total Internal Reflection in a well-polished rod to generate radially-polarized light. The use of Goos-Hanchen shift for polarization rotation allows ultra-broadband radial polarization conversion for wavelength bandwidth of >1 micron. In Ref. [24], we present a simple scheme that involves only standard off-the-shelf birefringent components to perform coherent summation of modes. This method is cheap, easy to implement as it does not require any special fabricating technique, and, it promises high power handling of more than kilowatts of laser power. Therefore it is a promising scheme for performing external radial polarization conversion for high power fiber laser.

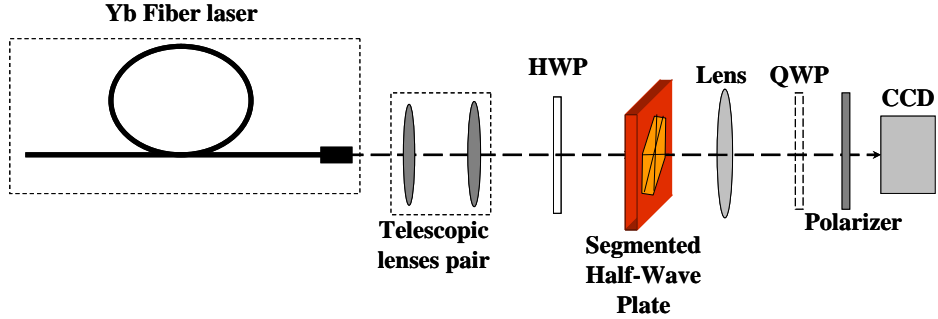

**Figure 1: Schematic of our experimental setup (HWP: Half-Wave-Plate and QWP: Quarter-Wave-Plate)**

In this paper, we investigate the performance of another promising high power external radial polarization converter. This time, it is based on a photonic crystal segmented half-wave plates. We have generated >100 Watts of average power of radial polarized beam from the linearly polarized near-gaussian beam of a Yb fiber laser. Unlike previous segmented half-wave-plate methods [1, 20] that use birefringence of bulk anisotropic crystals, this polarization converter utilizes the artificially engineered birefringence of photonics crystal [25, 26].

Figure 1 shows a schematic of our experimental setup. The lasers used are single transverse mode, linear polarized Yb fiber lasers. To demonstrate its power handling capability, we performed the radial polarization conversion using a single mode Yb fiber that can produce up to ~110 watts of linearly polarized 1064 nm laser output. This laser has a bandwidth of < 0.1nm, which is not a typical bandwidth (2-3nm) of high power fiber laser used in laser cutting and machining. Thus we also performed the radial polarization conversion using another Yb fiber laser that has a bandwidth of 2nm. This is to study whether these photonic crystal segmental half-wave-plates has an operating bandwidth acceptable by typical high power fiber laser. However, this Yb fiber laser only can produce up to ~14 watts of linearly polarized 1064 nm. In our experiment, the laser beam is expanded by a telescope lenses pair to a spot of ~5-6 mm to fill up the entire 7mm clear aperture of the segmental half-wave-plate. The near field beam after radial polarization conversion is imaged onto the CCD with and without the polarizer. To monitor the far field beam profile, we place the CCD camera at the focal point of a lens placed after the radial polarization converter. To measure the radial polarization purity, we capture four consecutive 2D beam profiles: 1) one observed with horizontal polarizer, 2) one observed with vertical polarizer 3) one observed with 45° and 4) finally, one observed with circular polarizer. A circular polarizer is basically a cascade of Quarter-Wave Plate (QWP) with a 45° slow axis and a horizontal polarizer. These four 2D beam profiles are then used to compute the Stokes vector distribution across the beam.

The photonic crystal segmented half-wave-plates used in our experiment is from Photonic Lattice, Inc. It consists of 12 segments of differently oriented 2D photonic crystals, as shown in Figure 2. The photonics crystal is fabricated using an auto-cloning technique described in

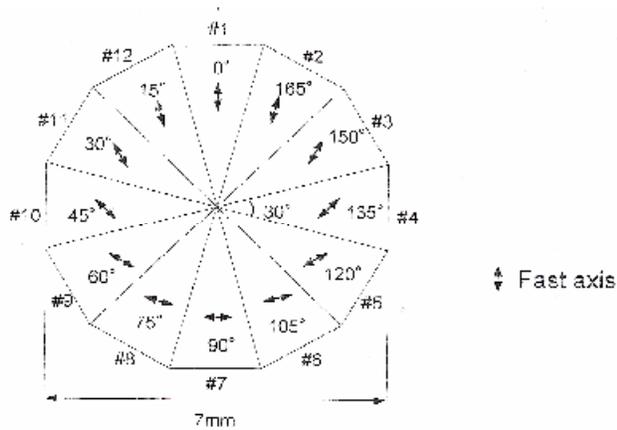

**Figure 2: Orientation of the fast axis of the twelve segments of Photonic Crystal Half-Wave-Plate**

Ref. [25] and the artificial birefringence is carefully engineered by controlling the number of lamination and the layer thickness of the photonics crystal [26]. To generate the radial polarized beam, the anisotropic axes of each segment have to be correctly oriented as shown in Figure 2. This is achieved by the initial patterning of grooves on the substrate which is followed by deposition of auto-cloned photonic crystal on the patterned substrate. By having an accurate control on the number of lamination and the layer thickness of the photonics crystal, the achieved retardations of our segmented half-wave-plate are measured to be within 0.1 radian variation from the desired half-wave retardation for all the twelve segments.

Theoretically, we calculated that, for this segmented half-wave-plate of 12 segments, the radial polarization purity can be as high as 99%, the calculated beam quality, $M^2$, of the radial polarized beam is 2.32 and the transformation efficiency calculated from the overlap integral with a perfect radial polarized $LG^*_{01}$ mode is 91%. This transformation efficiency is close to the theoretical limit of 93% of such segmental half-wave-plate conversion method [20].

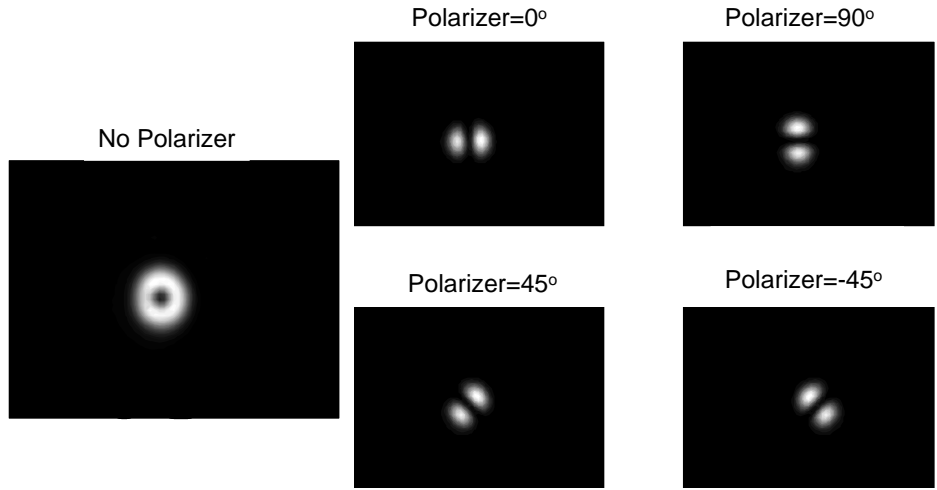

**Figure 3 : Far-field beam profiles of radial polarized beam observed with and without a polarizer**

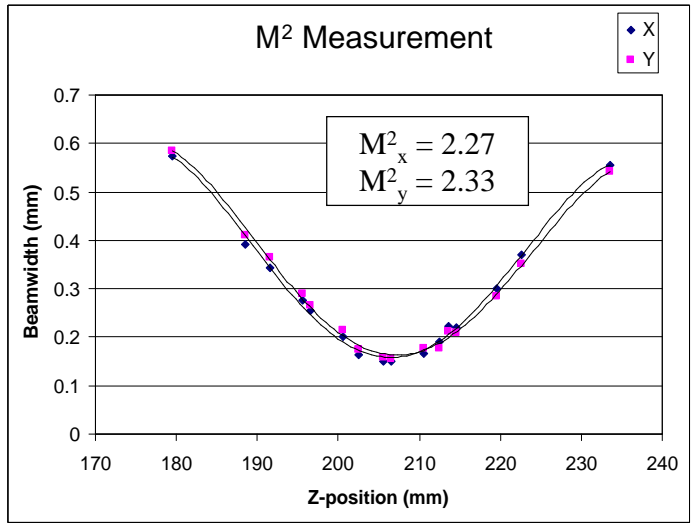

**Figure 4 : $M^2$ measurement of radial polarized beam**

In our experiment, we first used the 14 watts Yb fiber laser whose bandwidth is 2nm. Figure 3 shows the far field beam profiles of the radial polarized beam converted by the photonic crystal segmental wave-plate. They are observed with and without polarizer. Without a polarizer, the beam is a doughnut-shaped light beam as shown in Figure 3. When the polarizer is inserted prior to the CCD camera, two spots are clearly seen, and they rotate with the transmitting axis of the polarizer, as seen in Figure 3.

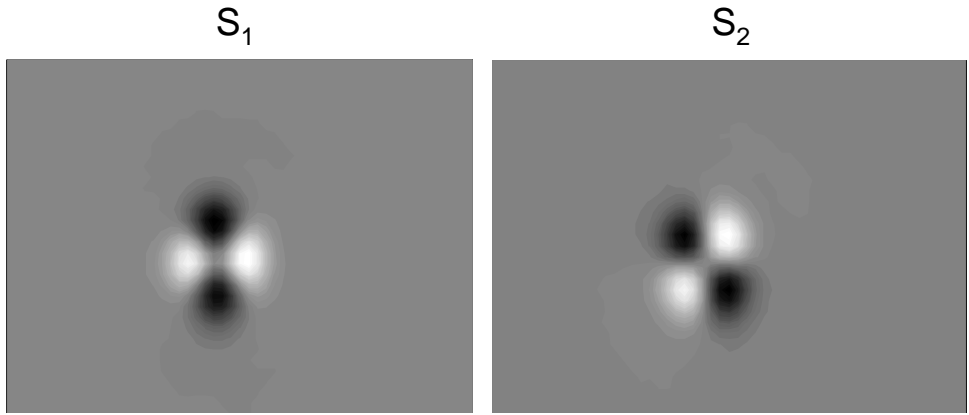

**Figure 5 : Distribution of $S_1$ and $S_2$ components of the Stokes vector across the beam**

For external radial polarization conversion method, due to the good quality of photonics crystal segmental half-wave-plate, there is hardly any power loss through transmission and the measured transmission is ~99.5%. Thus the generated radial polarized beam has a power of ~14 Watts. As shown in Figure 4, the beam quality in term of $M^2$ of the radial polarized beam is measured to be 2.27 and 2.32 in the x and y direction respectively. In addition, from our 2D polarimetric measurement, the radial polarization purity of the generated beam is measured to be ~98%. These measurements agree well with the theoretical value reported in earlier part of the paper. Figure 5 shows the measured $S_1$ and $S_2$ components of the Stoke vector distribution across the beam. The measured $S_3$ component of the Stoke vectors stays insignificant across the beam. This clearly shows the good radial purity of the beam and thereby illustrating that the photonics crystal segmented half-wave-plate has adequate operating bandwidth for typical high power fiber laser used in laser cutting industry. To demonstrate the high power handling, we change the laser to the 110 watts Yb laser of bandwidth < 0.1nm. With the high transmission of the photonics crystal segmental half-wave-plate, we observe negligible thermal effects and similar performance of the radial polarization conversion as before.

In conclusion, we have generated more than 100 watts of radial polarized beam from a Yb fiber laser using a segmented half-wave-plate. This method is promising for external radial polarization conversion of high power fiber laser used in the laser cutting industry.